\documentclass{webofc}
\usepackage[varg]{txfonts}   
\usepackage{amsmath}
\usepackage{amssymb}
\usepackage{epsfig}
\begin{document}
\title{How to measure the linear polarization of gluons in unpolarized proton  
using the heavy-quark pair production}
%
%

\author{\firstname{} \lastname{N. Ya. Ivanov}\inst{1,3}\fnsep\thanks{\email{nikiv@yerphi.am}} \and
    \firstname{} \lastname{A. V. Efremov}\inst{2}  \and
     \firstname{} \lastname{O. V. Teryaev}\inst{1,2}
}

\institute{Veksler and Baldin Laboratory of High Energy Physics, JINR, 141980 Dubna, Russia 
\and
           Bogoliubov Laboratory of Theoretical Physics, JINR, 141980 Dubna, Russia 
\and
           Yerevan Physics Institute, Alikhanian Br.~2, 0036 Yerevan, Armenia
          }

\abstract{%
In recent papers \cite{we_TMD,we_TMD_CG}, two new ways have been proposed to probe the linear polarization of gluons in unpolarized proton: using the azimuthal asymmetries and Callan-Gross ratio in heavy-quark pair leptoproduction, $lN\rightarrow l^{\prime}Q\bar{Q}X$. In this talk, we discuss in details the sensitivity of the QCD predictions for the azimuthal $\cos \varphi$ and $\cos 2\varphi$ asymmetries to the contribution of linearly polarized gluons inside unpolarized proton, where the azimuth $\varphi$ is the angle between the lepton scattering plane $(l,l^{\prime})$ and the heavy quark production plane $(N,Q)$. Our analysis shows that the azimuthal distributions under consideration vary from 0 to 1 depending on the transverse-momentum dependent gluonic counterpart of the Boer-Mulders function, $h_{1}^{\perp g}$. We conclude that the $\cos \varphi$ and $\cos 2\varphi$ asymmetries in heavy-quark pair production in DIS processes are predicted to be large in wide kinematic ranges and sensitive to the contribution of linearly polarized gluons.
}
\maketitle
\section{Introduction}
\label{intro}

Transverse momentum dependent (TMD) distributions of the transversely polarized quarks, $h_{1}^{\perp q}(\zeta,\vec{k}_{T}^2)$, and linearly polarized gluons, $h_{1}^{\perp g}(\zeta,\vec{k}_{T}^2)$, in an unpolarized nucleon play especial role in studies of the spin-orbit couplings of partons. Measurements of these quantities could clarify in details the proton spin decomposition puzzle. 

The $h_{1}^{\perp q}$ and  $h_{1}^{\perp g}$ densities have been introduced in Refs.~\cite{Boer-Mulders} and \cite{Mulders_2001}, respectively. Contrary to its quark counterpart (i.e. so-called Boer-Mulders function $h_{1}^{\perp q}$) the gluon TMD distribution, $h_{1}^{\perp g}$, is $T$- and chiral-even and thus can directly be probed in certain  electroproduction experiments.

In Refs.~\cite{Boer_HQ_1,Boer_HQ_2,Boer_HQ_3}, it was proposed to probe the $h_{1}^{\perp g}$ density using the azimuthal correlations in heavy-quark pair (and dijet) production in unpolarized electron–proton DIS. \footnote{The opportunities to measure the function $h_{1}^{\perp g}$ in unpolarized hadron-hadron collisions are discussed in Ref.~\cite{Boer_2015}.} The complete angular structure of the $Q\bar{Q}$ production cross section has been obtained in terms of seven azimuthal modulations. However, only two of these modulations are really independent \cite{we9}. In Ref.~\cite{we_TMD}, the leading order (LO) QCD predictions for the $\cos \varphi$ and $\cos 2\varphi$ distributions have been provided, where $\varphi$ is the heavy quark (or anti-quark) azimuthal angle. It was shown that these azimuthal asymmetries (AAs) are expected to be large in wide kinematic ranges and very sensitive to the function $h_{1}^{\perp g}$. For this reason, measurements of the AAs at the proposed EIC \cite{EIC} and LHeC \cite{LHeC2} colliders seem to be very promising for determination of the linear polarization of gluons inside unpolarized proton.

In the present talk, we discuss in details the pQCD predictions for the $\cos \varphi$ and $\cos 2\varphi$ asymmetries in charm and bottom pair leptoproduction. In particular, the $p_{\perp}$- and $z$\,-  distributions of the AAs at LHeC energies are considered.

\section{Production cross section}
\label{sec-1}
In Refs.~\cite{Boer_HQ_1,Boer_HQ_2,Boer_HQ_3}, the contribution of the linearly polarized gluons  to the reaction
\begin{equation} \label{1}
l(\ell )+N(P)\rightarrow l^{\prime}(\ell -q)+Q(p_{Q})+\bar{Q}(p_{\bar{Q}})+X(p_{X}) 
\end{equation}
with unpolarized initial states has been studied. To probe the TMD distributions, we need to measure (reconstruct) in the process (\ref{1}) the momenta of both heavy quark and anti-quark, $\vec{p}_{Q}$ and $\vec{p}_{\bar{Q}}$. To describe the contributions of TMD densities, the sum and difference of the transverse heavy quark momenta are used,
\begin{align} \label{3}
\vec{K}_{\perp}&=\frac{1}{2}(\vec{p}_{Q\perp}-\vec{p}_{\bar{Q}\perp}), &
\vec{q}_{T}&=\vec{p}_{Q\perp}+\vec{p}_{\bar{Q}\perp},
\end{align}
in the plane perpendicular to the direction of the target and the exchanged photon. The azimuthal angles of $\vec{K}_{\perp}$ and $\vec{q}_{T}$ are denoted by $\phi_{\perp}$ and $\phi_{T}$,  respectively. 

\begin{figure}[h]
\centering
\mbox{\epsfig{file=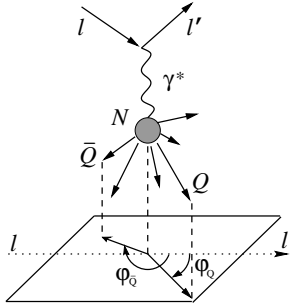,width=140pt}}
\caption{\label{Fg.1}\small Definition of the azimuthal angles $\varphi_Q$ and $\varphi_{\bar{Q}}$ in the nucleon rest frame.}
\end{figure}

Following Refs.~\cite{Boer_HQ_1,Boer_HQ_2,Boer_HQ_3}, we use the approximation when $\vec{q}_{T}^2\ll \vec{K}_{\perp}^2$, $\vec{p}^2_{Q\perp}\simeq \vec{p}^2_{\bar{Q}\perp}\simeq\vec{K}_{\perp}^2$, and the outgoing heavy quark and anti-quark are almost back-to-back in the transverse plane, see Fig.~\ref{Fg.1}. 

The contribution of the photon-gluon fusion mechanism to the reaction (\ref{1}) has the following factorized form:
\begin{equation} \label{8}
{\rm d}\sigma\propto L(\ell,q)\otimes \Phi_g(\zeta, k_{T})\otimes \left| H_{\gamma^*g\rightarrow Q\bar{Q}X} (q,k_{g},p_{Q},p_{\bar{Q}})\right|^2, 
\end{equation}
where $L^{\alpha\beta}(\ell,q)=-Q^2 g^{\alpha\beta}+2(\ell^{\alpha}\ell^{\prime\beta}+\ell^{\beta}\ell^{\prime\alpha})$ is the leptonic tensor and $H_{\gamma^*g\rightarrow Q\bar{Q}X}(q,k_{g},p_{Q},p_{\bar{Q}})$ is the amplitude for the hard partonic subprocess. The symbol $\otimes$ stands for appropriate convolutions: the phase space integration and traces over the color and Dirac indices.  

Information about parton densities in unpolarized nucleon is formally encoded in  corresponding TMD parton correlators. In particular, the gluon correlator is usually parameterized  as \cite{Mulders_2001}
\begin{equation} \label{9}
\Phi_g^{\mu\nu}(\zeta, k_{T})\propto - g_T^{\mu\nu}f_{1}^{g}\big(\zeta,\vec{k}_{T}^2\big)+\left(g_T^{\mu\nu}-2\frac{k_T^\mu k_T^\nu}{k_T^2}\right)\frac{\vec{k}_{T}^2}{2m^2_N}h_{1}^{\perp g}\big(\zeta,\vec{k}_{T}^2\big), 
\end{equation}   
where $m_N$ is the nucleon mass, 
\begin{align} \label{10}
g_T^{\mu\nu}&=g^{\mu\nu}- \frac{P^\mu n^\nu+P^\nu n^\mu}{P\cdot n},& n^\mu &=\frac{q^\mu+xP^\mu}{P\cdot q},& r&=\frac{\vec{k}_{T}^2 h_{1}^{\perp g}}{2m^2_N f_{1}^{g}} .
\end{align}
In Eq.~(\ref{10}), the tensor $-g_T^{\mu\nu}$ is (up to a factor) the density matrix of unpolarized gluons. The TMD distribution $h_{1}^{\perp g}\big(\zeta,\vec{k}_{T}^2\big)$ describes  the contribution of linearly polarized gluons. The degree of their linear polarization is determined by the quantity $r=\frac{\vec{k}_{T}^2 h_{1}^{\perp g}}{2m^2_N f_{1}^{g}}$. In particular, the gluons are completely polarized along the $\vec{k}_{T}$ direction at $r=1$, $\vec{k}_g=\zeta \vec{P}+\vec{k}_{T}$. 

The LO predictions for the azimuth-dependent cross section of the reaction (\ref{1}) can be written as \cite{we_TMD}
\begin{align} 
&\frac{{\rm d}^{6}\sigma}{{\rm d}y\,{\rm d}x\,{\rm d}z\,{\rm d}\vec{K}_{\perp}^2{\rm d}\vec{q}_{T}^2{\rm d}\varphi}=\frac{e_{Q}^{2}\alpha_{em}^2\alpha_{s}}{8\,\bar{S}^2}\frac{f_{1}^{g}(\zeta,\vec{q}_{T}^2)\hat{B}_2}{y^3 x\,\zeta z\,(1-z)}\Bigg\{\left[1+(1-y)^2 \right]\left(1-2r \frac{\hat{B}^h_2}{\hat{B}_2}\right)-y^2\frac{\hat{B}_L}{\hat{B}_2}\left(1-2r \frac{\hat{B}^h_L}{\hat{B}_L}\right) \nonumber \\
&+2(1-y)\frac{\hat{B}_A}{\hat{B}_2}\left(1-2r \frac{\hat{B}^h_A}{\hat{B}_A}\right)\cos2\varphi +(2-y)\sqrt{1-y}\frac{\hat{B}_I}{\hat{B}_2}\left(1-2r \frac{\hat{B}^h_I}{\hat{B}_I}\right)\cos\varphi\Bigg\}, \label{14}
\end{align}
where $e_Q$ is the heavy quark charge; $x$, $y$ and $Q^2$ are the usual Bjorken variables; $\bar{S}=2\, \ell\cdot P$, $z=\frac{p_{Q}\cdot P}{q\cdot P}$, $\zeta=\frac{q\cdot p_{Q}}{q\cdot P}$  and $\varphi$ is the heavy quark azimuth, $\varphi=\varphi_Q$.

The coefficients $\hat{B}_i$ $(i=2,L,A,I)$ originate from the contribution of unpolarized gluons,  while the quantities $\hat{B}^h_i$ are associated with the function $h_{1}^{\perp g}$. The LO results for $\hat{B}_i$ and $\hat{B}^h_i$ $(i=2,L,A,I)$ are presented in Ref.~\cite{we_TMD}.

\section{Azimuthal $\cos2\varphi$ asymmetry}
\label{sec-2}

Let us first consider the $\cos 2\varphi$ distribution in the case when the unpolarized gluons only contribute to the reaction (\ref{1}), i.e. for $r=0$. At $y\ll 1$ and fixed values of $Q^2$, the corresponding asymmetry is:
\begin{equation} \label{19}
A_{\cos2\varphi}(z,\vec{K}_{\perp}^2)\simeq \hat{B}_A\big/\hat{B}_2.
\end{equation} 

As shown in Ref.~\cite{we_TMD}, the function $A_{\cos2\varphi}(z,\vec{K}_{\perp}^2)$ has an extremum  at $z=1/2$ and $\vec{K}_{\perp}^2=m^2+Q^2/4$, where $m$ is the heavy-quark mass. This maximum value is: $A_{\cos2\varphi}(z=1/2,\vec{K}_{\perp}^2=m^2+Q^2/4)=\frac{1}{3}.$ This implies that the maximum value of the azimuthal $\cos2\varphi$ asymmetry is the same for both charm and bottom quarks at arbitrary values of $Q^2$.
\begin{figure}
\centering
\begin{tabular}{cc}
\mbox{\epsfig{file=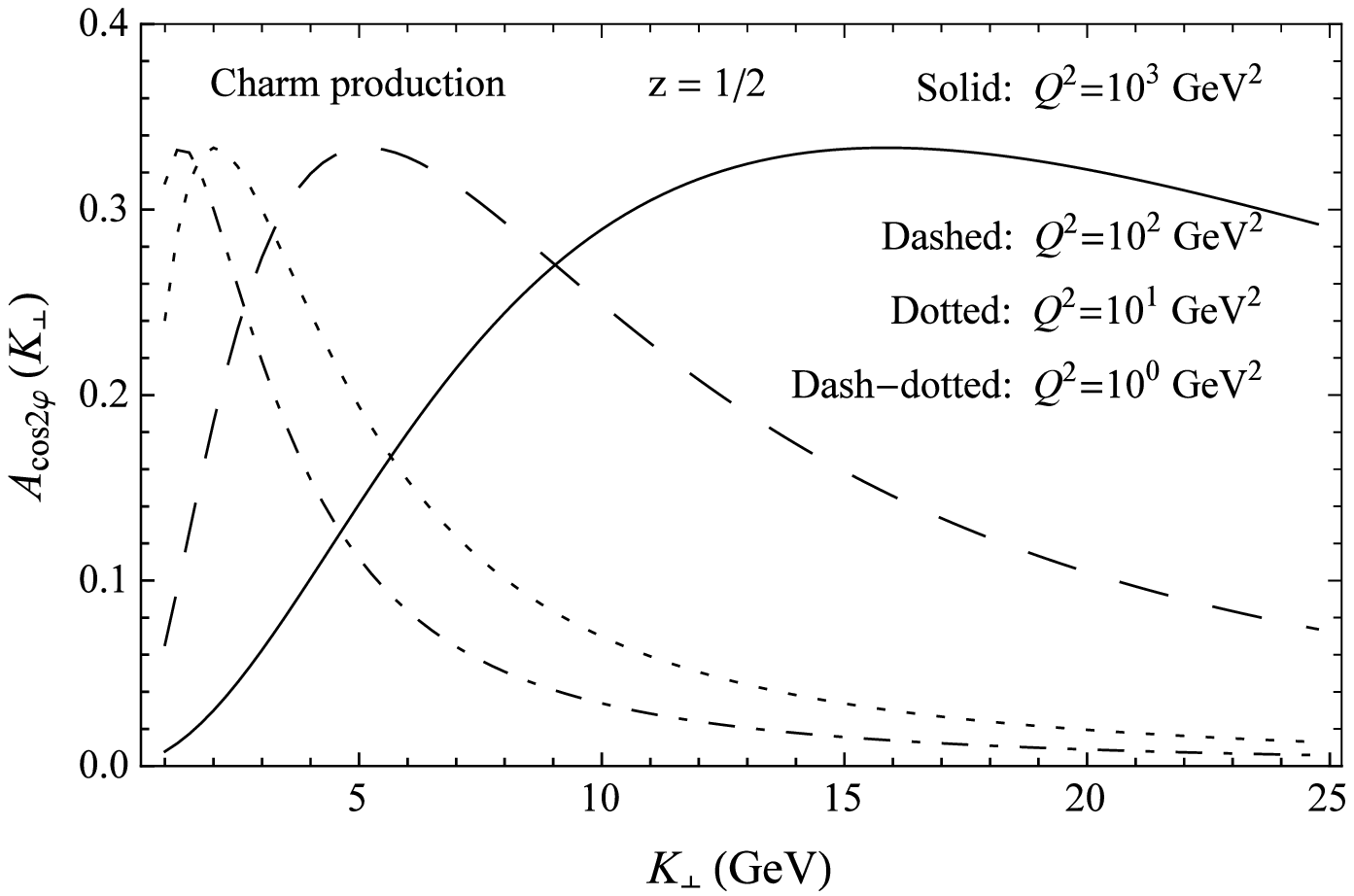,width=170pt}}
& \mbox{\epsfig{file=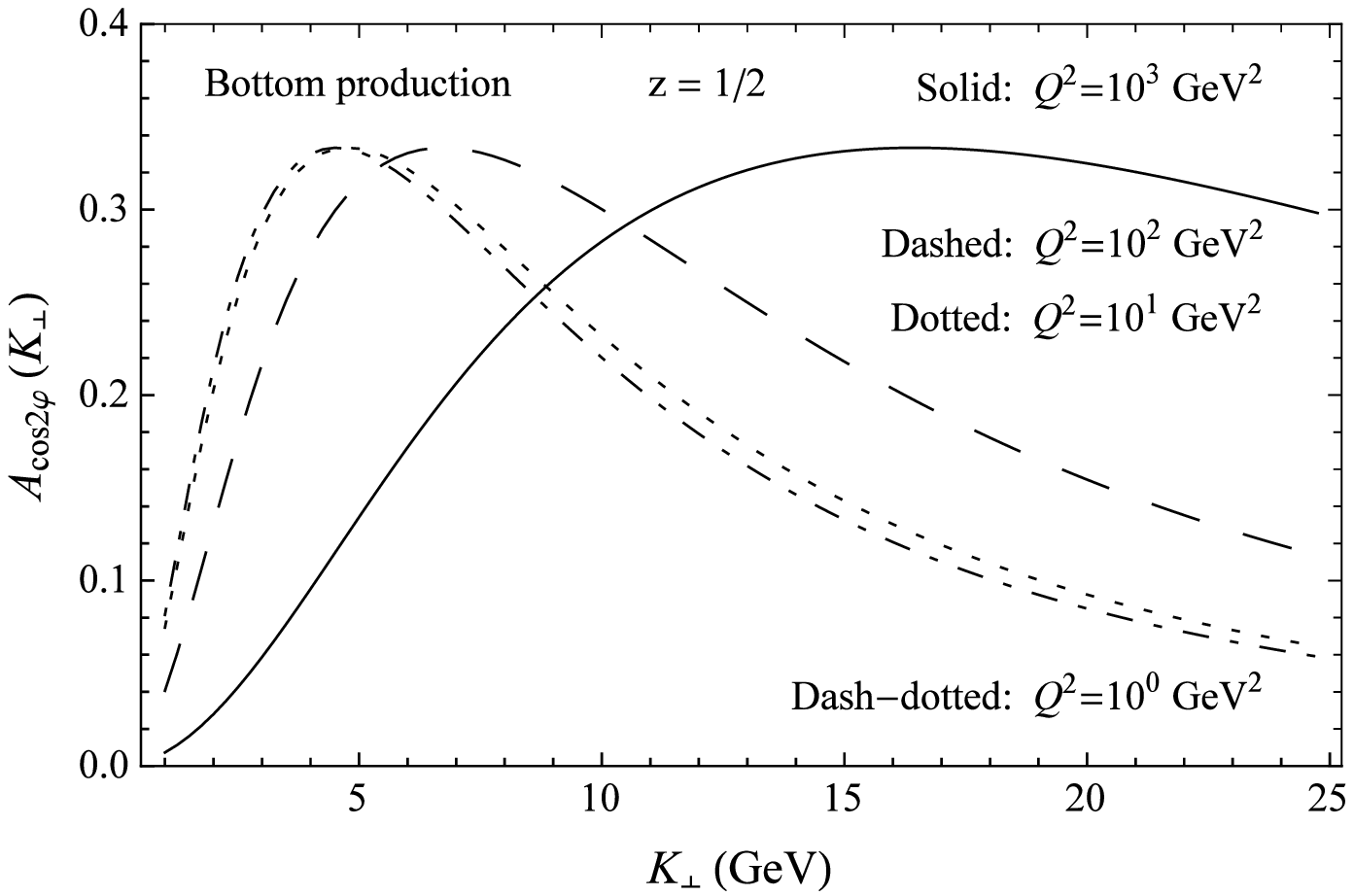,width=170pt}}\\
\end{tabular}
\caption{\label{Fg.2}\small Azimuthal $\cos2\varphi$ asymmetry $A_{\cos2\varphi}(K_{\perp})\equiv A_{\cos2\varphi}(z=1/2,K_{\perp})$ in charm ({\it left panel}) and bottom ({\it right panel}) production as a function of $K_{\perp}=\big|\vec{K}_{\perp}\big|$ at several values of $Q^2$.}
\end{figure}

The LO predictions for the asymmetry $A_{\cos2\varphi}(K_{\perp})\equiv A_{\cos2\varphi}(z=1/2,K_{\perp})$ in charm and bottom production as a function of $K_{\perp}=\big|\vec{K}_{\perp}\big|$ at several values of $Q^2$ are presented in Fig.~\ref{Fg.2}.\footnote{We use $m_c=$ 1.25 GeV and $m_b=$ 4.5 GeV.} One can see from Fig.~\ref{Fg.2} that sizable values for the $\cos2\varphi$ asymmetry are expected in wide regions of $K_{\perp}$ and $Q^2$.

\begin{figure}
\centering
\begin{tabular}{cc}
\mbox{\epsfig{file=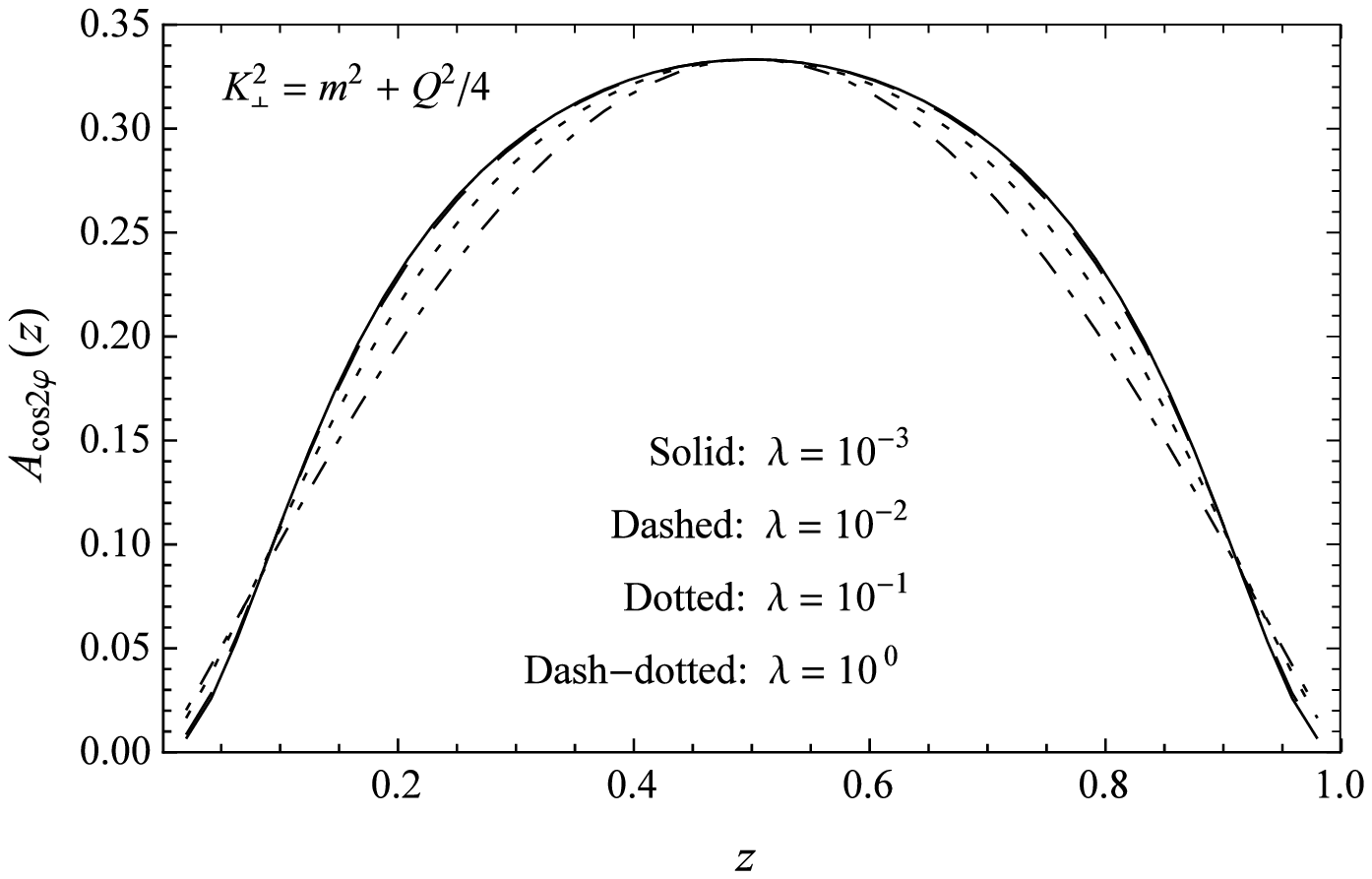,width=170pt}}
& \mbox{\epsfig{file=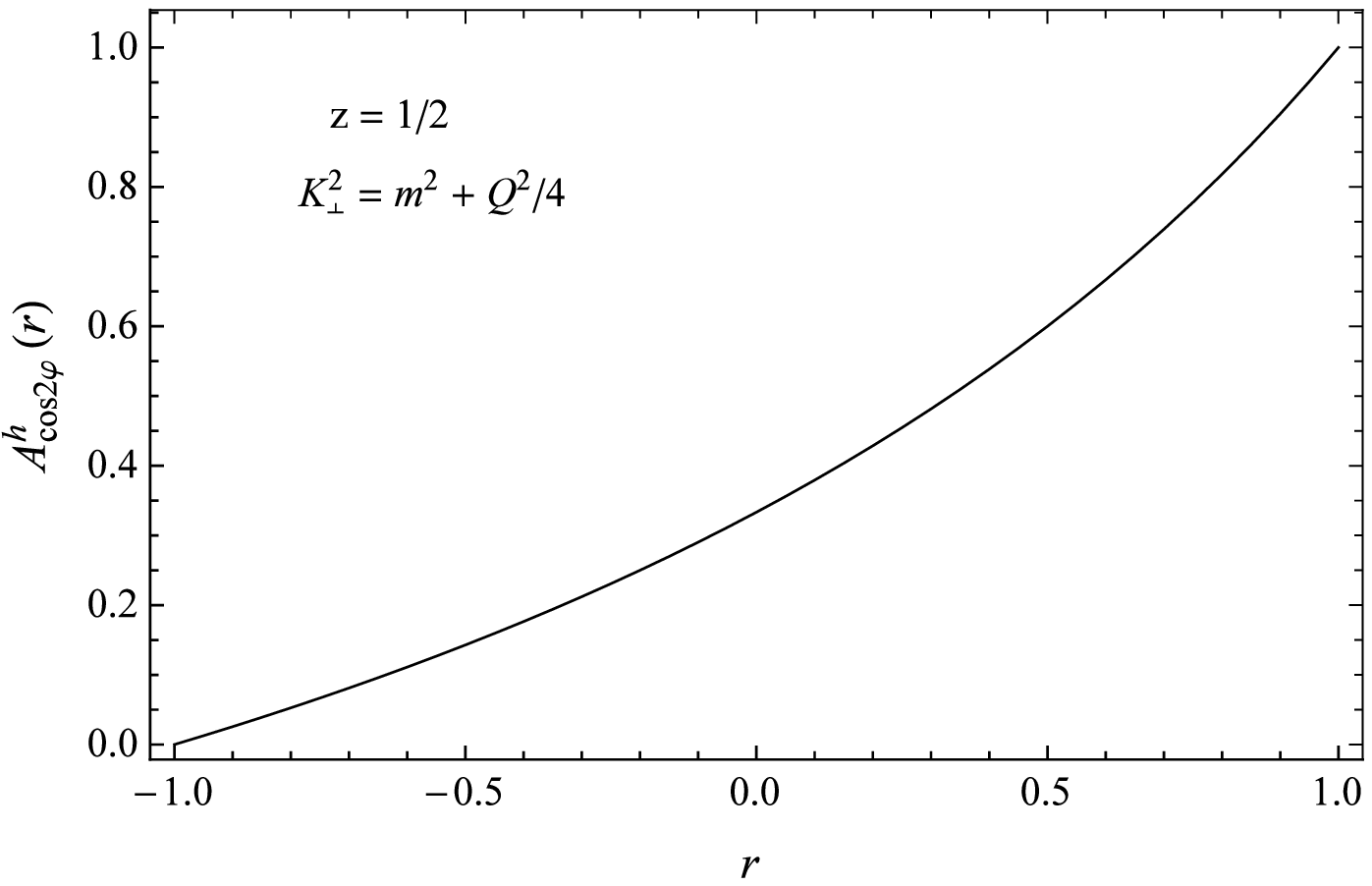,width=170pt}}\\
\end{tabular}
\caption{\label{Fg.3}\small {\it Left panel:} azimuthal $\cos2\varphi$ asymmetry $A_{\cos2\varphi}(z)\equiv A_{\cos2\varphi}(z,K_{\perp}^2=m^2+Q^2/4)$ in heavy-quark leptoproduction as a function of $z$ at several values of $\lambda=m^2\big/Q^2$. {\it Right panel:}  maximum value of the $\cos2\varphi$ asymmetry with the contribution of linearly polarized gluons, $A^h_{\cos2\varphi}(r)$, as a function of $r$.}
\end{figure}

The quantity $A_{\cos2\varphi}(z)\equiv A_{\cos2\varphi}(z,K_{\perp}^2=m^2+Q^2/4)$ in heavy-quark leptoproduction as a function of $z$ at several values of $\lambda=m^2\big/Q^2$ is presented in  Fig.~\ref{Fg.3} (left panel). One can see that this distribution is practically independent of $Q^2$ for both charm and bottom quarks.

Let us now discuss the contribution of the linearly polarized gluons to the $\cos 2\varphi$ distribution. In this case, the corresponding asymmetry, containing the contributions of both $f_1^g$ and $h_1^{\perp g}$ densities, is a function of three variables: 
\begin{equation} \label{21}
A^h_{\cos2\varphi}(z,\vec{K}_{\perp}^2,r)\simeq \frac{\hat{B}_A}{\hat{B}_2}\frac{1-2r\hat{B}^h_A\big/\hat{B}_A}{1-2r\hat{B}^h_2\big/\hat{B}_2}.
\end{equation}
Our analysis shows that the function $A^h_{\cos2\varphi}(z,\vec{K}_{\perp}^2,r)$ has a maximum at $z=1/2$ and $\vec{K}_{\perp}^2=m^2+Q^2/4$ for all values of $r$ in the interval $-1\leq r\leq 1$. We find \cite{we_TMD}
\begin{equation} \label{22}
A^h_{\cos2\varphi}(r)\equiv A^h_{\cos2\varphi}(z=1/2,\vec{K}_{\perp}^2=m^2+Q^2/4,r)=\frac{1+r}{3-r}.
\end{equation}
One can see from Eq.~(\ref{22}) that the maximum value of the $\cos2\varphi$ asymmetry with the contribution of linearly polarized gluons is a simple function of only variable $r$ (i.e. it is independent of $m^2$ and $Q^2$). The function $A^h_{\cos2\varphi}(r)$ is depicted in  Fig.~\ref{Fg.3} (right panel) where its strong dependence on the variable $r$ is seen.

\section{Azimuthal $\cos\varphi$ asymmetry}
\label{sec-3}

The azimuthal $\cos\varphi$ asymmetry due to the  contribution of unpolarized gluons only has the following form:  
\begin{equation} \label{23}
A_{\cos\varphi}(z,\vec{K}_{\perp}^2)\simeq\hat{B}_I\big/\hat{B}_2. 
\end{equation}  
Contrary to the $\cos2\varphi$ distribution, the quantity $A_{\cos\varphi}(z,\vec{K}_{\perp}^2)$ is an alternating function of both $z$ and $\vec{K}_{\perp}^2$. In particular, $A_{\cos\varphi}(z,\vec{K}_{\perp}^2)=-A_{\cos\varphi}(1-z,\vec{K}_{\perp}^2)$ and $\int{\rm d}z\, A_{\cos\varphi}(z,\vec{K}_{\perp}^2)=0$. 

Our analysis shows that the function (\ref{23}) has four extrema in the physical region of $z$ and $\vec{K}_{\perp}^2$: two maxima and two minima. We describe the extrema points in the ($z,\hat{k}^2$) plane with the help of four functions of $\lambda$: $z_{\pm}\equiv z_{\pm}(\lambda)$ and $\hat{k}^2_{\pm}\equiv\hat{k}^2_{\pm}(\lambda)$, where $\hat{k}^2\equiv\frac{K_{\perp}^2}{Q^2}$ and $\lambda\equiv\frac{m^2}{Q^2}$. The $\cos\varphi$ asymmetry takes its maximum and minimum values at ($z_{\pm},\hat{k}^2_{\pm}$) and ($z_{\pm},\hat{k}^2_{\mp}$), respectively. 

Exact results for the functions $z_{\pm}(\lambda)$ and $\hat{k}^2_{\pm}(\lambda)$ at arbitrary $\lambda$ are presented in Ref.~\cite{we_TMD}. At high $Q^2\gg m^2$, we have 
\begin{align}\label{27}
z_\pm(\lambda\rightarrow 0)&\simeq\genfrac{\{}{.}{0pt}{0}{\,0.841}{\,0.159},& \hat{k}^2_{\pm}(\lambda\rightarrow 0)&\simeq\genfrac{\{}{.}{0pt}{0}{\,0.707}{\,0.025}.
\end{align}

\begin{figure}
\centering
\begin{tabular}{cc}
\mbox{\epsfig{file=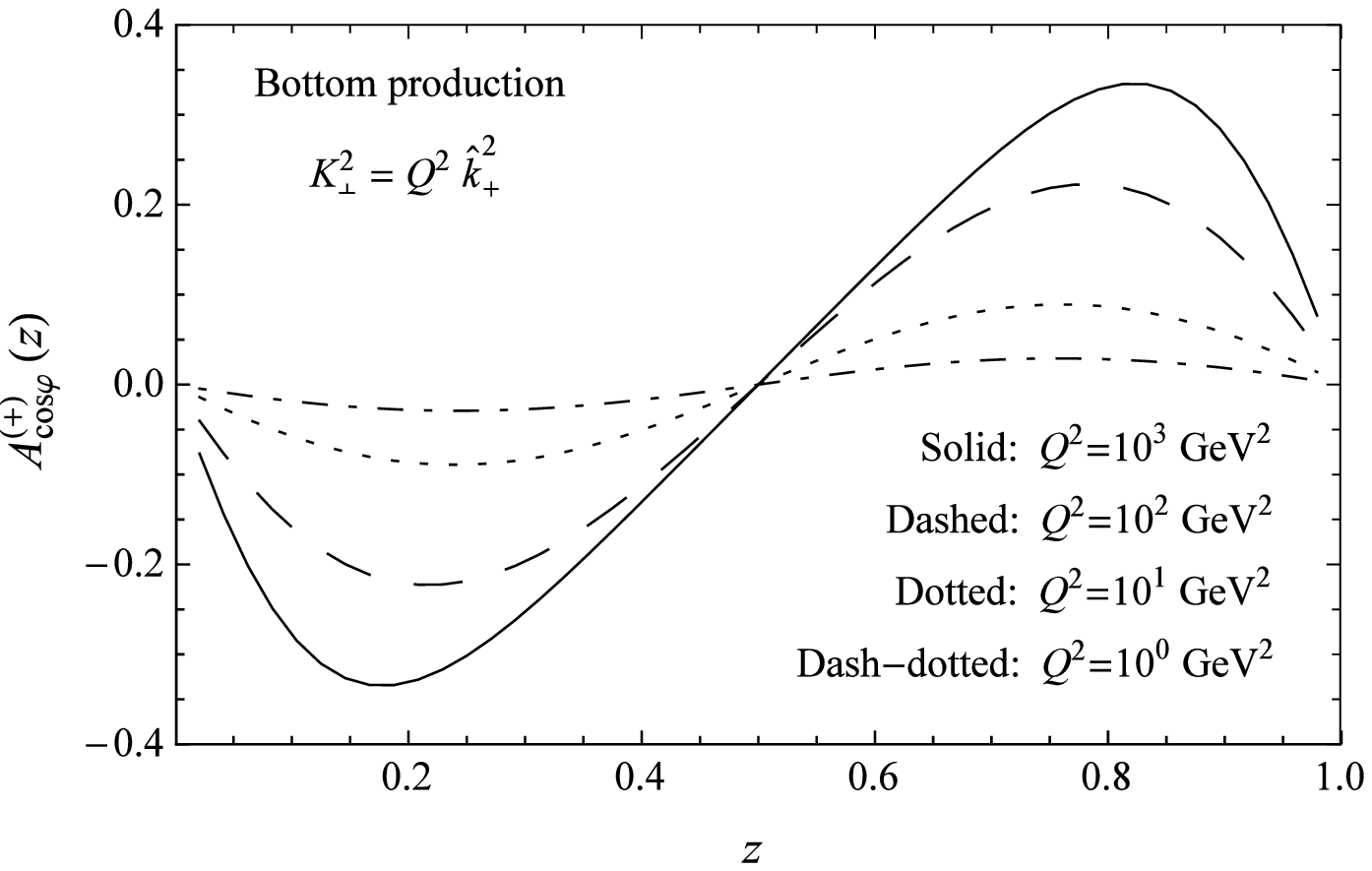,width=170pt}}
& \mbox{\epsfig{file=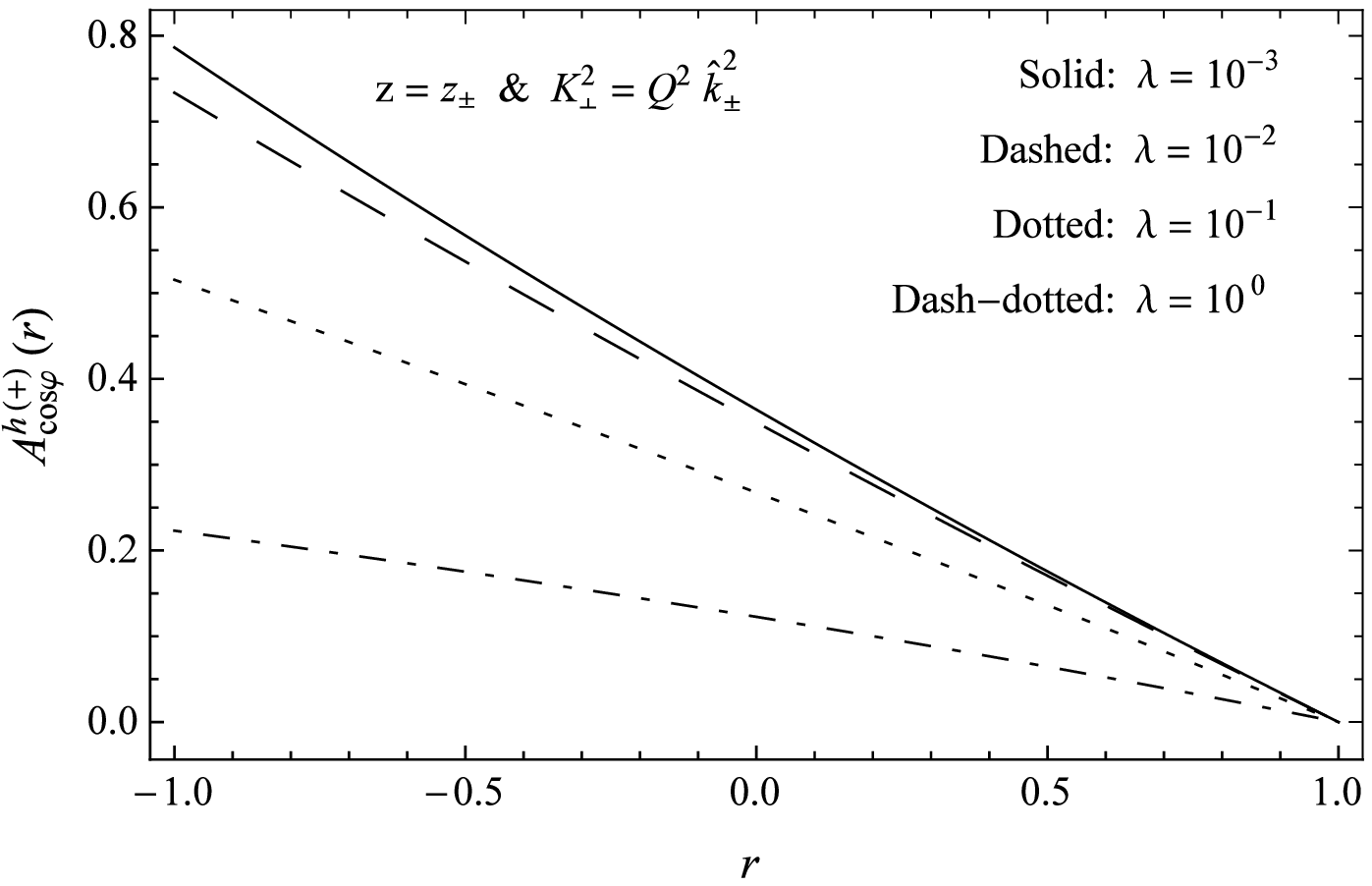,width=170pt}}\\
\end{tabular}
\caption{\label{Fg.6}\small {\it Left panel:} $\cos\varphi$ asymmetry due to the contribution of unpolarized gluons only, $A^{(+)}_{\cos\varphi}(z)\equiv A_{\cos\varphi}(z,K_{\perp}=Q^2\hat{k}^2_{+})$, as a function of $z$ at several values of $Q^2$. {\it Right panel:} $\cos\varphi$ asymmetry with the contribution of linearly polarized gluons, $A^{h(+)}_{\cos\varphi}(r)\equiv A^h_{\cos\varphi}(z=z_{\pm},\vec{K}_{\perp}^2=Q^2\hat{k}^2_{\pm},r)$, as a function of $r$ at several values of $\lambda$.}
\end{figure}

The $K_{\perp}$- behavior of the $\cos\varphi$ asymmetry, $A^{(\pm)}_{\cos\varphi}(K_{\perp})\equiv A_{\cos\varphi}(z=z_{\pm},K_{\perp})$, has been considered in details in Ref.~\cite{we_TMD}. In Fig.~\ref{Fg.6} (left panel), we present the quantity $A^{(+)}_{\cos\varphi}(z)\equiv A_{\cos\varphi}(z,K_{\perp}=Q^2\hat{k}^2_{+})$ as a function of $z$  at several values of $Q^2$. One can see that the maximum and minimum values of the asymmetry grow with $Q^2$.


At arbitrary $m^2$ and $Q^2$, the quantities $A_{\cos\varphi}(z=z_{\pm},\vec{K}_{\perp}^2=\hat{k}^2_{\pm}Q^2)$ are complicated functions of $\lambda$ which take their maximum and minimum values at  $\lambda\rightarrow 0$:
\begin{align}
A_{\cos\varphi}(z=z_{+},\vec{K}_{\perp}^2=Q^2\hat{k}^2_{\pm})&=-A_{\cos\varphi}(z=z_{-},\vec{K}_{\perp}^2=Q^2\hat{k}^2_{\pm})\stackrel{\lambda\rightarrow 0}{=}\pm\frac{\sqrt{3}-1}{2}\simeq \pm 0.366. \label{29} 
\end{align}
So, this value of the $\cos\varphi$ asymmetry, $(\!\sqrt{3}-1)/2$, is the maximal one allowed by the photon-gluon fusion with unpolarized initial gluons. 

Let us now consider the contribution of the linearly polarized gluons inside unpolarized nucleon to the $\cos\varphi$ distribution. The corresponding asymmetry, containing the contributions of both $f_1^g$ and $h_1^{\perp g}$ densities, is described by
\begin{equation} \label{30}
A^h_{\cos\varphi}(z,\vec{K}_{\perp}^2,r)\simeq \frac{\hat{B}_I}{\hat{B}_2}\frac{1-2r\hat{B}^h_I\big/\hat{B}_I}{1-2r\hat{B}^h_2\big/\hat{B}_2}.
\end{equation}
We denote the values of this function at $(z_\pm,\hat{k}^2_{\pm})$ and $(z_\pm,\hat{k}^2_{\mp})$ as $A^{h(+)}_{\cos\varphi}(r)$ and $A^{h(-)}_{\cos\varphi}(r)$, respectively:\footnote{At the same values of $\lambda$, $A^{h(-)}_{\cos\varphi}(r)=-A^{h(+)}_{\cos\varphi}(r)$. }
\begin{align} \label{31}
A^{h(+)}_{\cos\varphi}(r)&\equiv A^h_{\cos\varphi}(z=z_{\pm},\vec{K}_{\perp}^2=Q^2\hat{k}^2_{\pm},r),& A^{h(-)}_{\cos\varphi}(r)&\equiv A^h_{\cos\varphi}(z=z_{\pm},\vec{K}_{\perp}^2=Q^2\hat{k}^2_{\mp},r).
\end{align}
Contrary to the quantity $A^{h}_{\cos2\varphi}(r)$ given by Eq.~(\ref{22}), the functions $A^{h(\pm)}_{\cos\varphi}(r)$ depend explicitly on $\lambda$. For high $Q^2\gg m^2$, we have
\begin{align} \label{32}
A^{h(\pm)}_{\cos\varphi}(r)\stackrel{\lambda\rightarrow 0}{=}\pm \frac{\big(\!\sqrt{3}-1\big)\,\big(1-r\big)}{2-r\,\big(1-2\big/\sqrt{3}\big)}.
\end{align}

In Fig.~\ref{Fg.6} (right panel), the $\cos\varphi$ asymmetry with the contribution of linearly polarized gluons, $A^{h(+)}_{\cos\varphi}(r)\equiv A^h_{\cos\varphi}(z=z_{\pm},\vec{K}_{\perp}^2=Q^2\hat{k}^2_{\pm},r)$, is depicted as a function of $r$ at several values of $\lambda$. One can see from Fig.~\ref{Fg.6}  strong dependence of the asymmetry on both $r$ and $\lambda$. In particular, the upper bound on the absolute value of the function $A^{h(+)}_{\cos\varphi}(r)$ is equal to $\frac{2(\!\sqrt{3}-1)}{3-2/\sqrt{3}}\simeq  0.793$ (the case of $r\rightarrow-1$ and $\lambda\rightarrow 0$), while the lower bound vanishes (the case of $r\rightarrow 1$). 

\section{Conclusion}
\label{sec-4}

Our main conclusion is that both the $\cos\varphi$ and $\cos2\varphi$ asymmetries in heavy-quark pair leptoproduction could be good probes of the linear polarization of gluons inside unpolarized nucleon. First, these asymmetries are predicted to be large within pQCD: maximal values allowed by  the photon-gluon fusion with unpolarized gluons (i.e. when $h_{1}^{\perp g}$=\,0) are $(\!\!\sqrt{\,3}-1)/2$ and $1/3$, respectively. Second, the azimuthal distributions are very sensitive to the linear polarization of gluons: maximum values of both $\cos \varphi$ and $\cos 2\varphi$ asymmetries vary from 0 to 1 depending on $h_{1}^{\perp g}$.


Unfortunately, the AAs in heavy-quark electroproduction are presently unmeasured. At the same time, the $\cos2\varphi$ asymmetry is well defined in 1PI kinematics within pQCD: it is stable both perturbatively and parametrically \cite{we2,we4}. For this reason, it seems to be good probe of the heavy-quark densities \cite{we5,we6} (both intrinsic \cite{BHPS} and perturbative \cite{Collins}) and linearly polarized gluon distribution, $h_{1}^{\perp g}$, in unpolarized proton \cite{we_TMD}.

{\it Acknowledgements.} The authors are grateful to S.~J.~Brodsky and A.~Kotzinian for useful discussions. This work is supported in part by the ANSEF grant PS-nuclth-5027.   

\def\baselinestretch{.9}

\end{document}